\documentclass[aps,showpacs,preprintnumbers,amsmath,amssymb]{revtex4}
\usepackage{dcolumn}
\usepackage{multirow}
\usepackage[dvips]{epsfig}
\def \be {\begin{equation}}
\def \ee {\end{equation}}
\def \bea {\begin{eqnarray}}
\def \eea {\end{eqnarray}}

\begin{document}
\baselineskip=0.8 cm
\title{\bf Chaotic motion of particles in the accelerating and rotating black holes spacetime}
\author{Songbai Chen$^{1}$\footnote{Corresponding author: csb3752@hunnu.edu.cn}, Mingzhi Wang, Jiliang
Jing$^{1}$ \footnote{jljing@hunnu.edu.cn}}

\affiliation{Institute of Physics and Department of Physics, Hunan
Normal University,  Changsha, Hunan 410081, People's Republic of
China \\ Key Laboratory of Low Dimensional Quantum Structures \\
and Quantum Control of Ministry of Education, Hunan Normal
University, Changsha, Hunan 410081, People's Republic of China\\
Synergetic Innovation Center for Quantum Effects and Applications,
Hunan Normal University, Changsha, Hunan 410081, People's Republic
of China}

\begin{abstract}
\baselineskip=0.6 cm
\begin{center}
{\bf Abstract}
\end{center}

We have investigated the motion of timelike particles along
geodesic in the background of accelerating and rotating black
hole spacetime. We confirm that chaos exists in the geodesic
motion of the particles by Poincar\'e sections, the frequency spectrum and the power spectrum, the fast Lyapunov exponent indicator and the bifurcation diagram. Moreover, we probe the effects of the acceleration and rotation
parameters  on the chaotic behavior of a timelike
geodesic particle in the  black hole spacetime. Our results show that the acceleration  brings richer physics for the geodesic motion of particles.

\end{abstract}

\maketitle
\newpage

\section{Introduction}

Chaos is a kind of seeming random, chance or irregular motion, which appears only in the nonlinear and non-integrable dynamical systems. The
most important feature of chaos is that it is highly sensitive to initial conditions \cite{Sprott,Ott,Brown1}. The tiny errors in the chaotic motion can grow rapidly with time so that the motion is totally changed from what it would be in the absence of these errors, which means that it is very difficult to make a long-term prediction for chaotic motions in general. Thus, it is expected that chaotic systems possess many novel properties not shared by the usual dynamical systems, which triggers more attention to focus on the study of chaotic dynamics in various physical fields.

It is well known that the geodesic motion of particle in the generic Kerr-Newman black hole spacetime \cite{Carter} is integrable and chaos does not emerge in this system. In order to study the chaotic motions of particles in general relativity, one must resort to some spacetimes with complicated geometries or introduce some extra interactions to ensure that the dynamical system describing motion of particle is non-integrable.
Along this spirit, Cornish \textit{et al} \cite{Cornish,Hanan} investigated the chaotic trajectories of particle in multi-black hole spacetimes \cite{MP} where the equations of motion of particle are not variable-separable. Moreover, the chaotic motions of particles have been studied in the perturbed Schwarzschild spacetime \cite{Bombelli,Bombelli1,Bombelli2,Bombelli3}, or in the spacetime of a black hole immersed in magnetic field \cite{Karas}, or in the non-standard Kerr black hole spacetime described by Manko-Novikov metric \cite{Contopoulos,Contopoulos1,Contopoulos2,Contopoulos3}. The chaotic phenomenon was also found for the charged particles moving in a magnetic field interacting with gravitational waves \cite{HD}. More interestingly, Frolov and Larsen \cite{Frolov} showed that after introducing ring strings instead of point particles, one can find that the ring string dynamics is chaotic even in the asymptotically flat Schwarzschild black hole spacetime. Subsequently, the chaotic behavior of the ring string is also found in AdS-Schwarzschild black hole \cite{Zayas} and AdS-Gauss-Bonnet black hole spacetimes \cite{MDZ} .

In this paper, we will investigate the chaotic motion of  particle in
the accelerating and rotating black holes
spacetime\cite{Hong.k2005,J.B.Griffiths2005, J.B.Griffiths2006},
which describes two uniformly separated Kerr-type black holes accelerating away from each other under the action of `` strings "
represented by conical singularities located along appropriate
sections of the axis of symmetry. Comparing with the usual Kerr black hole, the accelerating and rotating black
holes spacetime has different geometric structure because it possesses two rotation horizons and two acceleration horizons. This implies that both Hawking and Unruh radiation could be present
in this background. Saifullah \textit{et al} \cite{JM2,US2} studied the
surface gravity, Hawking temperature and the area laws for
accelerating and rotating black holes and explored the effects
of the acceleration of black holes on the Hawking radiation of scalar particles in these black holes spacetime. We \cite{Yao} investigated the collision of two massive geodesic particles in the accelerating and rotating black hole spacetime and probe the properties of the
center-of-mass energy and high-velocity collision belts in the near
horizon collision. Moreover, with this accelerating and rotating metric, Hawking and Ross \cite{Hawking} researched
the possibility that a black hole pair can be created by the breaking of a
cosmic string. Since the study of such a kind of
black holes could provide physical insight into the high energy
physics, the properties of the accelerating and
rotating black holes have been investigated extensively in recent years
\cite{Hong.k2005,J.B.Griffiths2005, J.B.Griffiths2006, JM2,US2, Yao, Hawking,
JH1}. The main purpose of this paper is to investigate the chaotic dynamics in the accelerating and rotating black holes spacetime.

The paper is organized as follows. In Sec. II, we briefly review the
accelerating and rotating black hole spacetime and discuss the equations of
geodesic motion for a timelike particle in this background. In Sec. III, we
investigate the chaotic phenomenon in the accelerating and
rotating black holes spacetime. We end the paper
with a summary.

\section{Geodesic motion of a particle in the
accelerating and rotating black hole spacetime}

The accelerating and rotating black hole spacetime describes the gravitational field by a pair of uniformly accelerating Kerr-type black holes, which is a special case of the Pleba\'{n}ski and Demia\'{n}ski metric \cite{JM1}
covered a large family of electro-vacuum type-$D$ spacetimes
including both the Kerr-Newman like solutions and the $C$-metric. In the Boyer-Lindquist
coordinates, the metric of this accelerating and rotating black
holes spacetime has a form
\cite{Hong.k2005,J.B.Griffiths2005, J.B.Griffiths2006}
\begin{eqnarray}
ds^{2} &=&-\left( \frac{\Delta-a^{2}P\sin ^{2}\theta }{\rho ^{2}\Omega
^{2}} \right) dt^{2}+\left( \frac{\rho ^{2}}{\Delta\Omega ^{2}}\right)
dr^{2}+\left(
\frac{\rho ^{2}}{P\Omega ^{2}}\right) d\theta ^{2}  \nonumber \\
&&+\left( \frac{\sin ^{2}\theta \left[ P\left( r^{2}+a^{2}\right)
^{2}-a^{2}\Delta\sin ^{2}\theta \right] }{\rho ^{2}\Omega ^{2}}\right)
d\phi ^{2}
\nonumber \\
&&-\left( \frac{2a\sin ^{2}\theta \left[ P\left( r^{2}+a^{2}\right)
-\Delta\right] }{\rho ^{2}\Omega ^{2}}\right) dtd\phi, \label{metric0}
\end{eqnarray}
with
\begin{eqnarray}
\Omega &=&1-\alpha r\cos \theta , \label{1.1} \\
\rho ^{2} &=&r^{2}+a^{2}\cos ^{2}\theta ,  \label{1.2} \\
P &=&1-2\alpha M\cos \theta + \alpha ^{2}
a^{2}\cos ^{2}\theta ,  \label{1.3} \\
\Delta &=&(r^{2}-2Mr+a^{2})(1-\alpha ^{2}r^{2}) . \label{1.4}
\end{eqnarray}
Here the parameters $M$, $\alpha$ and $a$ denote the mass, the acceleration and the angular momentum per unit mass of the black hole, respectively. The locations of black hole horizons are determined by equation $g^{rr}=0$
\cite{Hong.k2005,J.B.Griffiths2005, J.B.Griffiths2006}, i.e.,
\begin{eqnarray}
\frac{\Delta \Omega^2}{\rho^2}=0.
\end{eqnarray}
Solving above eqaution, one can obtain
\begin{equation}
 r_{H }=M+ \sqrt{M^{2}-a^{2}},\;\;\;\;\;r_{C}=M-\sqrt{M^{2}-a^{2}},\;\;\;\;\;
 r_A=\frac{1}{\alpha },\;\;\;\;r_{\alpha}=\frac{1}{\alpha \cos\theta}. \label{3}
\end{equation}
It is easy to find that the position of the event horizon $r=r_H$ and Cauchy
horizon $r=r_{C}$ are same to those of the Kerr black hole. However,
in this case, one can find that there also exist other two horizons at $r_A=\frac{1}{\alpha
}$ and $r_{\alpha}=\frac{1}{\alpha \cos\theta}$, which are interpreted as the acceleration horizons in the context of the $C$-metric. Obviously, the presence of acceleration changes the geometry of spacetime. Unlike in the usual Kerr black hole spacetime, the physical region of the black hole is located in $r_H<r<r_A$ in
which $\Delta>0$ is satisfied.

In the curve spacetime, the Lagrangian of a timelike particle moving
along the geodesic is
\begin{equation}
\mathcal{L}=\frac{1}{2}g_{\mu\nu}\dot{x}^{\mu}\dot{x}^{\nu},
\end{equation}
where the dots denote derivatives with respect to the proper time
$\tau$. For the
accelerating and rotating black hole spacetime
(\ref{metric0}), the Lagrangian takes the form
\begin{eqnarray}
\mathcal{L}&=&\frac{1}{2} \bigg\{-\bigg(\frac{\Delta-a^{2}P\sin ^{2}\theta }{\rho ^{2}\Omega
^{2}}\bigg)\dot{t}^2+ \frac{\rho ^{2}\dot{r}^2}{\Delta\Omega ^{2}}+\frac{\rho ^{2}\dot{\theta}^2}{P\Omega ^{2}}
-\left( \frac{2a\sin ^{2}\theta \left[ P\left( r^{2}+a^{2}\right)
-\Delta\right] }{\rho ^{2}\Omega ^{2}}\right)\dot{t}\dot{\varphi}\nonumber\\&+&\left( \frac{\sin ^{2}\theta \left[ P\left( r^{2}+a^{2}\right)
^{2}-a^{2}\Delta\sin ^{2}\theta \right] }{\rho ^{2}\Omega ^{2}}\right)\dot{\varphi}^2\bigg\}.
\end{eqnarray}
Making use of the Euler-Lagrangian equation, we obtain the equations
of motion of the timelike particles
\begin{eqnarray}
\dot{t}&=&\frac{g_{33}E+g_{03}L}{g^2_{03}-g_{00}g_{33}}=\frac{\Omega^{2}}{\Delta
P\rho^2} \bigg\{[P\left( r^{2}+a^{2}\right) ^{2}-a^{2}\Delta\sin
^{2}\theta]E-a[ P\left( r^{2}+a^{2}\right)
-\Delta]L\bigg\},\label{10}
\end{eqnarray}
\begin{eqnarray}
\dot{\varphi}&=&-\frac{g_{03}E+g_{00}L}{g^2_{03}-g_{00}g_{33}}=\frac{\Omega^{2}}{\Delta
P\rho^2\sin^2\theta} \bigg\{a\sin^2\theta[ P\left(
r^{2}+a^{2}\right)
-\Delta]E+a[\Delta-a^2P\sin ^{2}\theta]L\bigg\},\label{11}
\end{eqnarray}
\begin{eqnarray}
\ddot{r}&=&\frac{\Delta}{2\rho^2}
\bigg\{\bigg[\frac{2(\Delta-a^2P\sin^2\theta)}{ \rho^2}
\bigg(\frac{\Omega_{,r}}{\Omega}+\frac{\rho_{,r}}{\rho}\bigg)-
\frac{\Delta_{,r}-a^2P_{,r}\sin^2\theta}{\rho^2}\bigg]\dot{t}^2
+\frac{\rho^2}{
\Delta}\bigg[\frac{\Delta_{,r}}{\Delta}+2\bigg(\frac{\Omega_{,r}}{\Omega}
-\frac{\rho_{,r}}{\rho}\bigg)\bigg]\dot{r}^2\nonumber\\
&+&\frac{\rho^2}{P}\bigg[2\bigg(
\frac{\rho_{,r}}{\rho}-\frac{\Omega_{,r}}{\Omega}\bigg)-\frac{P_{,r}}{P}\bigg]\dot{\theta}^2
+\frac{4\rho^2}{\Delta}\bigg[\frac{\Omega_{,\theta}}{\Omega}
-\frac{\rho_{,\theta}}{\rho}\bigg]\dot{r}\dot{\theta}\nonumber\\&-&
\frac{2a\sin ^{2}\theta \left[ P\left( r^{2}+a^{2}\right)
-\Delta\right]
}{\rho^{2}}\bigg[\frac{P_{,r}(r^2+a^2)+2rP-\Delta_{,r}}{P\left(
r^{2}+a^{2}\right)
-\Delta}-2\bigg(\frac{\Omega_{,r}}{\Omega}+\frac{\rho_{,r}}{\rho}\bigg)\bigg]\dot{t}\dot{\varphi}
\nonumber\\&+&\frac{\sin ^{2}\theta \left[ P\left(
r^{2}+a^{2}\right) ^{2}-a^{2}\Delta\sin ^{2}\theta \right] }{\rho
^{2}}\bigg[\frac{P_{,r}(r^{2}+a^{2})^2+2r(r^2+a^2)P-a^{2}\Delta_{,r}\sin
^{2}\theta}{P\left( r^{2}+a^{2}\right) ^{2}-a^{2}\Delta\sin
^{2}\theta}-2\bigg(\frac{\Omega_{,r}}{\Omega}+\frac{\rho_{,r}}{\rho}\bigg)\bigg]\dot{\varphi}^2\bigg\},
\nonumber\\ \label{eqmotionr}
\end{eqnarray}
\begin{eqnarray}
\ddot{\theta}&=&\frac{P}{2\rho^2}
\bigg\{\bigg[\frac{2(\Delta-a^2P\sin^2\theta)}{ \rho^2}
\bigg(\frac{\Omega_{,\theta}}{\Omega}+\frac{\rho_{,\theta}}{\rho}\bigg)+
\frac{a^2\sin\theta(P_{,\theta}\sin\theta+2P\cos\theta)}{\rho^2}\bigg]\dot{t}^2
+\frac{2\rho^2}{ \Delta}\bigg[\frac{\Omega_{,\theta}}{\Omega}
-\frac{\rho_{,\theta}}{\rho}\bigg]\dot{r}^2\nonumber\\
&+&\frac{\rho^2}{P}\bigg[\frac{P_{,\theta}}{P}-2\bigg(
\frac{\rho_{,\theta}}{\rho}-\frac{\Omega_{,\theta}}{\Omega}\bigg)\bigg]\dot{\theta}^2
+\frac{\rho^2}{P}\bigg[2\bigg(
\frac{\rho_{,r}}{\rho}-\frac{\Omega_{,r}}{\Omega}\bigg)-\frac{P_{,r}}{P}\bigg]\dot{r}\dot{\theta}\nonumber\\&-&
\frac{2a\sin ^{2}\theta \left[ P\left( r^{2}+a^{2}\right)
-\Delta\right] }{\rho^{2}}\bigg[\frac{P_{,\theta}(r^2+a^2)}{P\left(
r^{2}+a^{2}\right)
-\Delta}-2\bigg(\frac{\Omega_{,\theta}}{\Omega}+\frac{\rho_{,\theta}}{\rho}
-\frac{\cos\theta}{\sin\theta}\bigg)\bigg]\dot{t}\dot{\varphi}
\nonumber\\&+&\frac{\sin ^{2}\theta \left[ P\left(
r^{2}+a^{2}\right) ^{2}-a^{2}\Delta\sin ^{2}\theta \right] }{\rho
^{2}}\bigg[\frac{P_{,\theta}(r^{2}+a^{2})^2-2a^{2}\Delta\sin
\theta\cos\theta}{P\left( r^{2}+a^{2}\right) ^{2}-a^{2}\Delta\sin
^{2}\theta}-2\bigg(\frac{\Omega_{,\theta}}{\Omega}+\frac{\rho_{,\theta}}{\rho}
-\frac{\cos\theta}{\sin\theta}\bigg)\bigg]\dot{\varphi}^2\bigg\},\nonumber\\
\label{eqmotion}
\end{eqnarray}
with the constraint condition
\begin{eqnarray}
H_1=P\dot{r}^2+\Delta\dot{\theta}^2-\frac{\Omega^4}{\rho^4\sin^2\theta}\bigg\{P\sin^2\theta[(r^2+a^2)E-aL]^2
-\Delta[a\sin^2\theta E-L]^2\bigg\}+ \frac{\Delta
P\Omega^2}{\rho^2}=0.\label{eqmotion1}
\end{eqnarray}
Here $E$ and $L$ correspond to the energy and angular momentum of
the timelike particle, respectively. It is obvious that in the case
with the non-zero acceleration, i.e., $\alpha \neq 0$, the equations
of motion (\ref{eqmotionr}), (\ref{eqmotion}) and (\ref{eqmotion1}) can not be variable-separable and the
corresponding dynamical system is non-integrable
 because it admits only two integrals of motion $E$ and $L$, which
implies that the motion of the particle could be chaotic in the
four-dimensional accelerating and rotating black hole spacetime
(\ref{metric0}).

\section{Chaotic phenomenon in the accelerating and
rotating black hole spacetime}

Chaos is a class of very complex motion without accurate definition
at present. Usually, it can be understood as a kind of seeming
random, chance or irregular movement appeared in a definiteness
system with nonlinear interaction and it is very sensitive to initial value.
The chaotic phenomenon in dynamical systems can be
detected by many kinds of methods including the Poincar\'{e}
surfaces of section, the Lyapunov characteristic exponents, the fast Lyapunov indicators (FLI), the power spectrum, the fractal basin boundaries, the bifurcation diagram, and so
on.
\begin{figure}[ht]
\includegraphics[width=15cm]{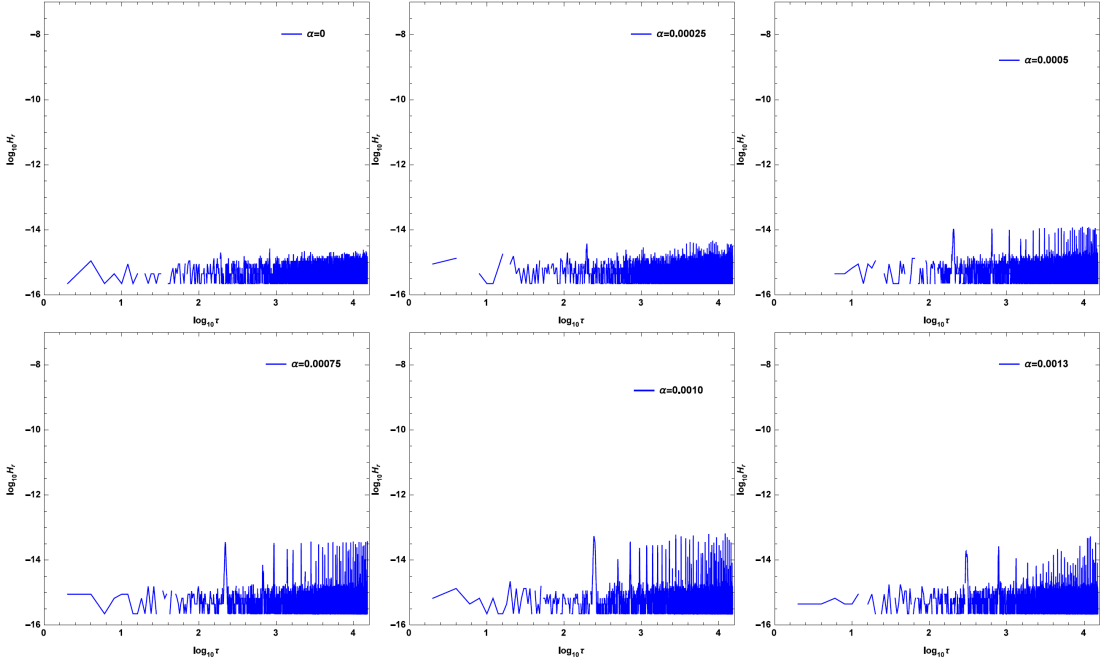}
\caption{Errors of $H_1$ with time computed by
the velocity correction method (RK5+Correction) in the accelerating
and rotating black hole spacetime for different $\alpha$. Here, we set
the parameters $E=0.95$, $L=3M$, $a=0.6$, and the initial
conditions $\{$ $r(0)=8$; $\dot{r}(0)=0.1$; $\theta(0)
=\frac{\pi}{2}$$\}$.}\label{f0}
\end{figure}

In order to investigate the dynamical properties of a chaotic system
with some coupled and complicated differential equations, we must
resort to the numerical method with high precision because the
motion of a particle in chaotic region is very sensitive to initial
value and the numerical errors may produce a big difference between
the numerical behavior and the real motion of particle.  Here, we
adopt to the corrected fifth-order Runge-Kutta method suggested in
literatures \cite{Wu1,Wu2}, in which the velocities $(\dot{r}, \dot{\theta})$ are corrected in integration and the numerical deviation is pulled back in a least-squares shortest path. As in refs.\cite{Wu1,Wu2}, the energy of the dynamical system (\ref{eqmotion}) is subjected to the constraint $H_1=0$, which means that $H_1$ could be regarded as a conserved quantity. However, the numerical errors in the integral calculation could yield some deviations so that the numerical solution $(\dot{r}, \dot{\theta}, r, \theta)$ does not satisfy the constraint $H_1=0$. In order to solve this problem, one can introduce a dimensionless parameter $\xi$ to make a connection between the numerical velocities $(\dot{r}, \dot{\theta})$ and the true value $(\dot{r}^{*}, \dot{\theta}^{*})$ in the form of
\begin{eqnarray}
\dot{r}^{*}=\xi\dot{r},\;\;\;\;\;\;\;\dot{\theta}^{*}=\xi\dot{\theta}.\label{scal1}
\end{eqnarray}
The scale $\xi$ can be chosen such that the constraint (\ref{eqmotion1}) is always satisfied. Inserting Eq.(\ref{scal1}) into Eq. (\ref{eqmotion1}), one can find that the scale factor of
velocity correction $\xi$ in the accelerating and
rotating black hole spacetime (\ref{metric0}) is
\begin{eqnarray}
\xi&=&\sqrt{\frac{\frac{\Omega^4}{\rho^4\sin^2\theta}\bigg\{P\sin^2\theta[(r^2+a^2)E-aL]^2-\Delta
[a\sin^2\theta E-L]^2\bigg\}- \frac{\Delta
P\Omega^2}{\rho^2}}{P\dot{r}^2+\Delta\dot{\theta}^2}}.
\end{eqnarray}
In this way, the precision of the conserved quantity $H_1$ in the system of Eqs. (\ref{10})-(\ref{eqmotion1}) at every integration step can hold perfectly. In Fig.(1), we present the change of $H_1$ with time computed by
the velocity correction method (RK5+Correction) in the accelerating
and rotating black hole spacetime for different $\alpha$. Here, we set
the parameters $E=0.95$, $L=3M$, $a=0.6$, and the initial
conditions $\{$ $r(0)=8$; $\dot{r}(0)=0.1$; $\theta(0)
=\frac{\pi}{2}$$\}$. From Fig.(1), one can find that the value of $H_1$ is
remained below $10^{-13}$ for different values of $\alpha$ and then the error is controlled greatly, which
displays sufficiently that this correction method is very powerful
so that it can avoid the pseudo chaos caused by numerical errors.

With help of the corrected fifth-order Runge-Kutta method, we
present some solutions for $r(\tau)$ for different values of $\alpha$ in Fig.(2), which are obtained under the rotation
parameter $a=0.6$ and the set of initial conditions $\{$ $r(0)=8$;
$\dot{r}(0)=0.1$; $\theta(0) =\frac{\pi}{2}$$\}$.
\begin{figure}[ht]
\includegraphics[width=15cm]{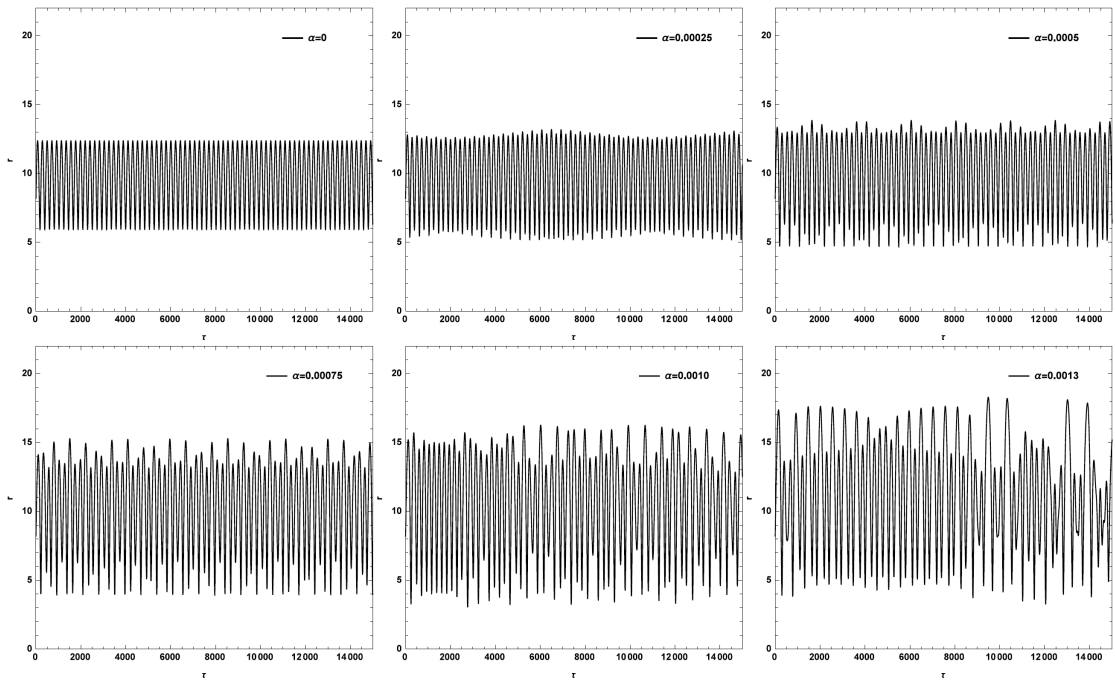}
\caption{ The change of $r(\tau)$ with $\tau$ for different $\alpha$ in
the accelerating and rotating black hole spacetime with
the parameters $E=0.95$, $L=3M$, and $a=0.6$. The initial
conditions are set by $\{$ $r(0)=8$; $\dot{r}(0)=0.1$; $\theta(0)
=\frac{\pi}{2}$$\}$. }\label{f1}
\end{figure}
\begin{figure}
\includegraphics[width=15cm]{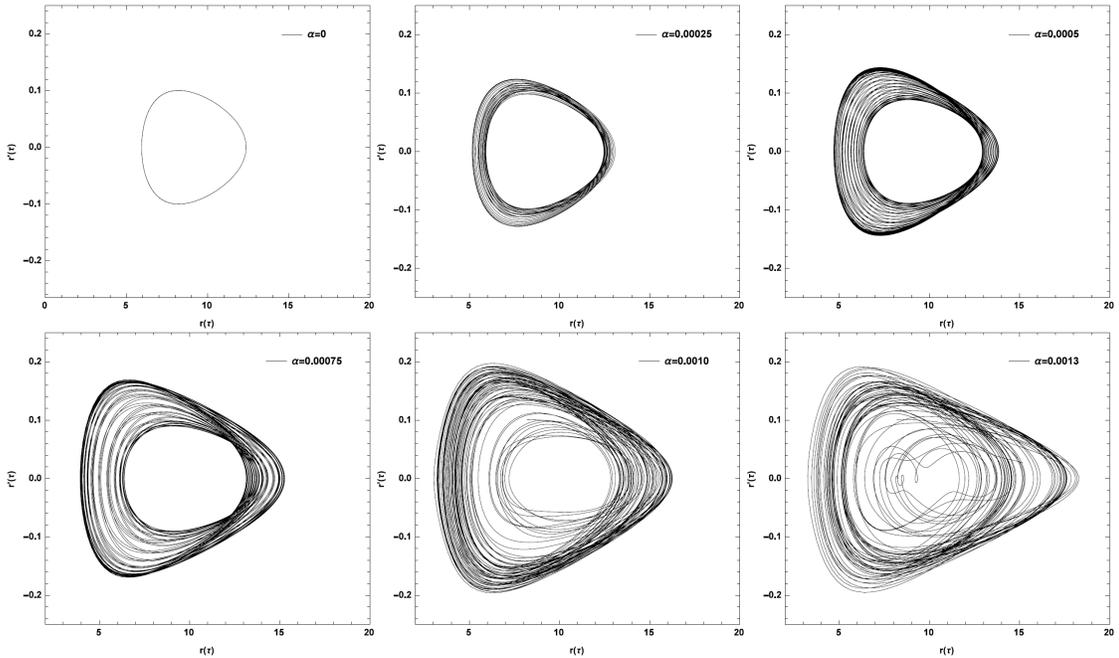}
\caption{ Phase curves corresponding to the solutions plotted in figure (\ref{f1}),
projected into the $(r, \dot{r})$ plane. }
\end{figure}
It is shown that the solution with $\alpha=0$ presented in Fig.(2)
is a periodic solution and is not chaotic, which is explained by a
fact that in the case $\alpha=0$ the metric (\ref{metric0}) reduces
to the usual Kerr black hole spacetime in which the timelike
geodesic equations are variable-separable and the chaos does not emerge in
such an integrable dynamical system.
However, for the cases with $\alpha\neq0$, we find that the amplitude and frequency components increase with $\alpha$. Especially, as $\alpha\geq0.001$,  it is difficult to describe the
amplitudes and frequencies of this oscillation with any definite
pattern, which implies that the motion could be chaotic. Thus, the presence of acceleration make the motion of particle more complicated. We also plot the phase curve in the $(r, \dot{r})$ plane
of the phase space for this
trajectory in Fig.(3). It is shown that the phase path in the case with $\alpha=0$ is simple and
is only a closed curve, which means that the corresponding solution
is periodic and the particle moves along the stable periodic orbit
around the black hole. However, with the increase of $\alpha$, the phase path becomes more complex and the region fulled by the path is enlarged, which means that the degree of disorder and non-integrability of the motion of particle increases with the acceleration parameter $\alpha$ in a sense for the signals plotted in Fig.(2). For the cases with $\alpha\geq0.001$, one can find that the complex path fulls densely a given region in the phase plane. It is a typical feature of chaotic behaviors in the definiteness systems and then the chaotic motion occurs in these cases. For the case $\alpha>0.0016$, we obtain only a kind of unstable escaped solutions for the chosen initial conditions, which describe that the particle falls into the acceleration horizon or the event horizon after undergoing several chaotic oscillations around black hole. This could be explained by a fact that for the larger $\alpha$ the acceleration horizon $r_A$ is closer to the event horizon $r_H$ so that the particle is out-off-balance in the physical region around the black hole ($r_H<r<r_A$). Thus, we will focus on the cases with $0\leq\alpha<0.0016$ in this paper.
\begin{figure}[ht]
\includegraphics[width=15cm]{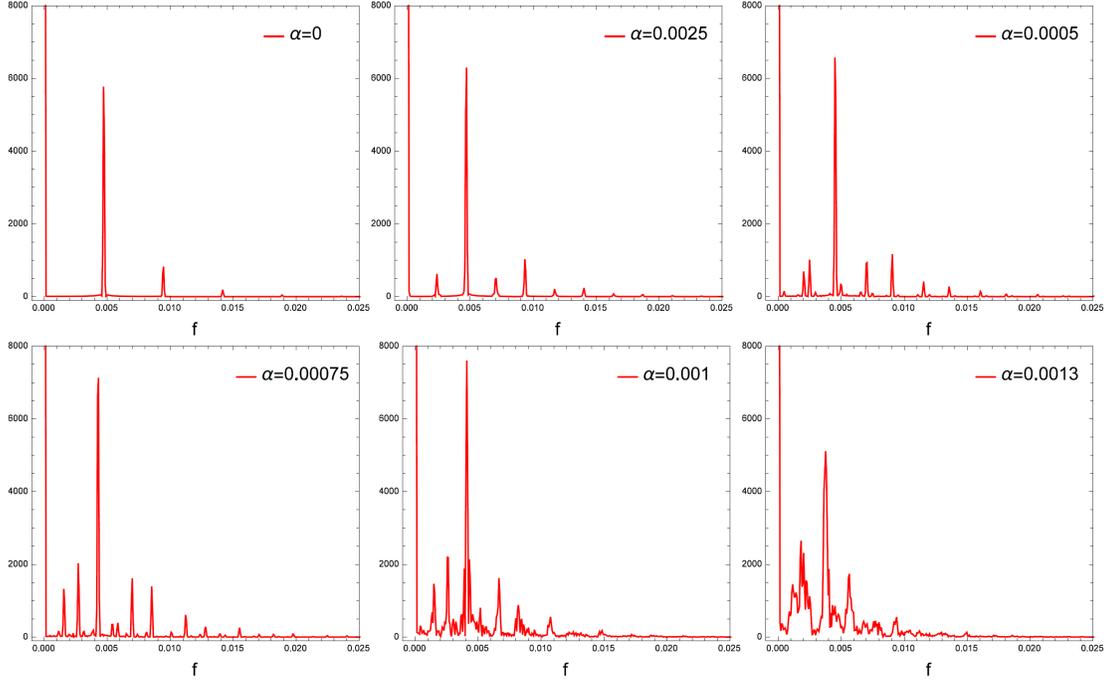}
\caption{ The frequency spectrum for the signals with different $\alpha$ plotted in figure (\ref{f1}). }\label{f1s}
\end{figure}
\begin{figure}[ht]
\includegraphics[width=15cm]{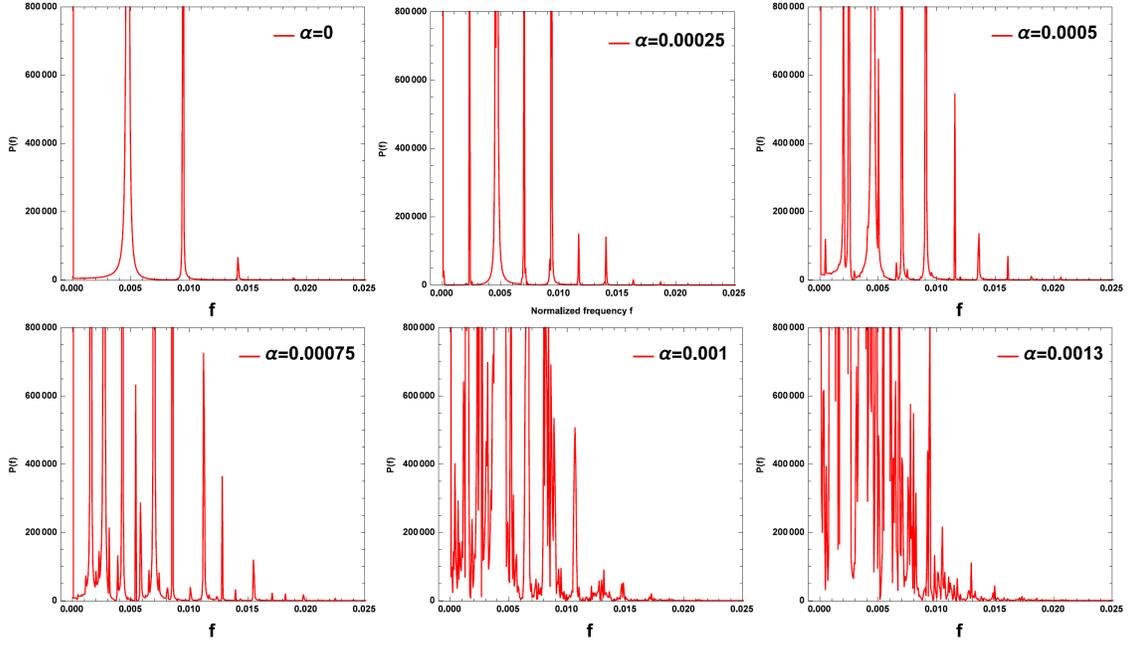}
\caption{ The power spectrum for the signals with different $\alpha$ plotted in figure (\ref{f1}).}
\end{figure}

Let us now to analyze the frequency components of the signals with different values of $\alpha$ plotted in Fig.(2). Through the fast Fourier transformation, we present the spectrogram
for the above signals in Fig.(4). The height of peaks are related to the amplitude of the corresponding frequency
in a Fourier decomposition \cite{Sprott,Ott}. In Fig.(4), one can find that there exists a high peak at the frequency $f=0$ Hz in each panel, whose amplitude describes the average orbital radius of the particle moving around the black hole. For the signal with $\alpha=0$ in Fig.(2), we also find that the peaks appear at the points with the frequency $f=f_0$, $2f_0$, $3f_0$ and $4f_0$, respectively. Here $f_0$ is the lowest frequency and its numerical value is $0.004772$. It is a discrete spectrum, which means that in the case $\alpha=0$ the motion of particle is multiple-periodic motion rather than a simple single periodic motion for chosen parameters and initial conditions. However, comparing with the amplitude of fundamental frequency part, the amplitudes of the overtone frequency parts are very small, which yields that the phase path looks like a close curve in the $r-\dot{r}$ plane as in the case of single periodic motion. Fig.(4) tells us that with the increase of $\alpha$, the frequency components increase and spectral lines become more dense. As $\alpha=0.001$, we find that the distinct continuous spectrum appears, which means that the motion of particle is chaotic in this case. As $\alpha=0.0013$, the width of continuous spectrum increases and the strength of chaotic motion enhances. The similar results are also obtained by analyzing the power spectrum shown in Fig.(5).

The Lyapunov indicator is a kind of useful tools to identify whether the motion of particles are chaotic or not \cite{Sprott,Ly1,Ly2,Ly3}. It is well known that the behavior of dynamical system is chaotic if the largest Lyapunov exponent is positive and is non-chaotic if the largest Lyapunov exponent is negative. In general, in order to obtain the Lyapunov exponent, one has to spend a long computational time even for a chaotic orbit since it is an infinite-time quantity.  However,  it has been shown that the relevant information on distinguishing between regular and chaotic trajectories could be obtained by integrating the equations of motion for a short time \cite{Ly4}.  FLI is such a kind of faster and more sensitive indicators to reveal chaos orbits of particles. Froeschl\'{e} and Lega \cite{Ly5} describe the FLI as $FLI(\tau)=\ln|Y(\tau)|$, where $Y(\tau)$ is a tangential vector of the flow at time $\tau$ as the particle moves along the trajectory. For a continuous flow along the trajectory including the periodic orbits, there exists a differential rotation which induces that the angle between the vectors of an initial orthonormal basis decreases sharply \cite{Ly4,Ly5,Ly6}. Due to the conservation of  volume for the continuous flow in the phase space, the differential rotation leads to a rapid increase of the tangential vector $Y(\tau)$  of the flow \cite{Ly4,Ly5,Ly6}, which means that FLI blows up for late times. The investigation \cite{Ly4,Ly5,Ly6} show that FLI($\tau$) grows with exponential rate for chaotic motion, even for weak chaotic motion, and  grows algebraically with time for the regular resonant orbit and for the periodic one.
Since the deviation vector $\Delta X$ between two nearby trajectories can approximate well the tangent vector, FLI can also be simplified as $FLI(\tau)=\ln\frac{|\Delta X(\tau)|}{|\Delta X(0)|}$.
This is so-called the two-particle method
 or two-nearby-trajectories method \cite{Ly7,Ly8}. It is a less rigorous but
still useful technique. The version of
FLI with two-nearby-trajectories in general relativity is described as \cite{MDZ,Wu5}
\begin{eqnarray}
FLI(\tau)=-k[1+\log_{10}d(0)]+\log_{10}\bigg|\frac{d(\tau)}{d(0)}\bigg|,
\end{eqnarray}
where $d(\tau)=\sqrt{|g_{\mu\nu}\Delta x^{\mu}\Delta x^{\nu}}|$,
$\Delta x^{\mu}$ is the deviation vector  between two nearby
trajectories at proper time $\tau$.  The parameter $k$ is the
sequential number of renormalization which is used to avoid that the
two orbits expand too fast.
In Fig.(6), we present FLI($\tau$) for the
signals plotted in  Fig.(2). It is shown that with increase of time $\tau$, FLI($\tau$) grows with
exponential rate for the signals with $\alpha=0.001$ and $\alpha=0.0013$, but
with polynomial rate in the cases with  $\alpha<0.001$. This confirms further that in Fig.(2) the orbital in with $\alpha=0.001$ and $\alpha=0.0013 $ are chaotic and the orbital with $\alpha<0.001$ is
ordered.
\begin{figure}[ht]
\includegraphics[width=15cm ]{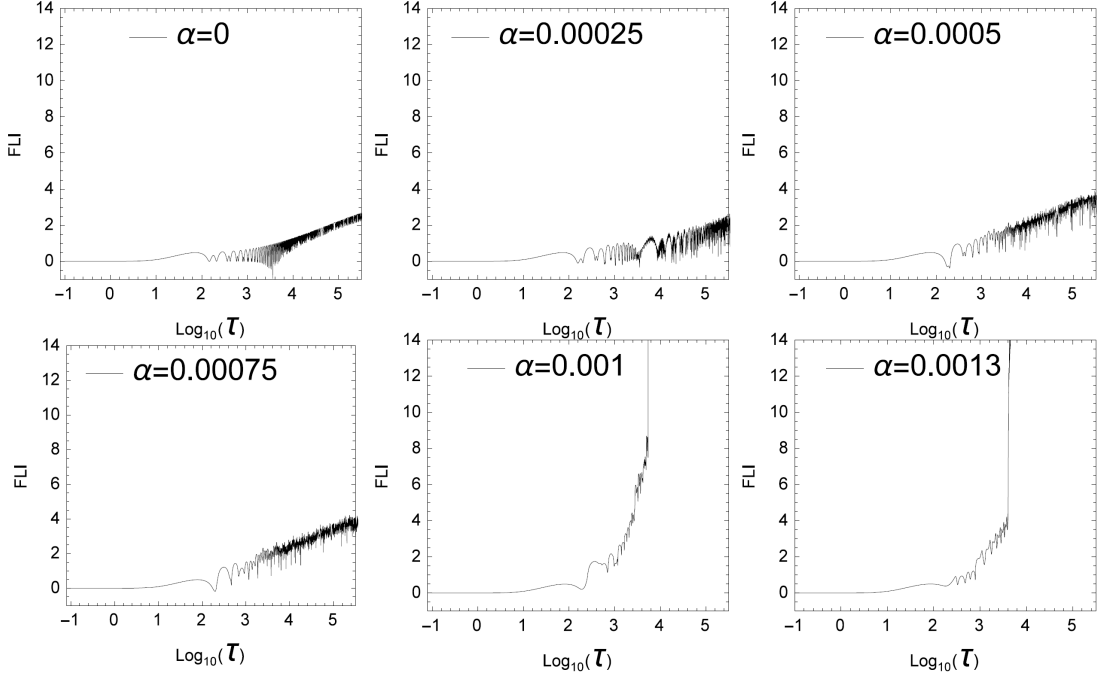}
\caption{ The Fast Lyapunov indicators with two nearby trajectories
for the solutions with different $\alpha$ plotted in figure (\ref{f1}).}
\end{figure}

Poincar\'{e} section is another useful tool for analyzing dynamical
systems. It is defined as an intersection of trajectory of a
continuous dynamical system with a given hypersurface which is
transversal to the trajectory in the phase space. In general, the
solutions of the continuous dynamical system with different initial conditions can be classified as
three kinds by the intersection points in a Poincar\'{e} section.
One of them are periodic solutions, which are corresponds to a finite
number of points in the Poincar\'{e} section. The quasi-periodic
solutions correspond to a series of close curves and the chaos
solutions correspond to strange patterns of dispersed points with
complex boundaries.
\begin{figure}[ht]
\includegraphics[width=15cm]{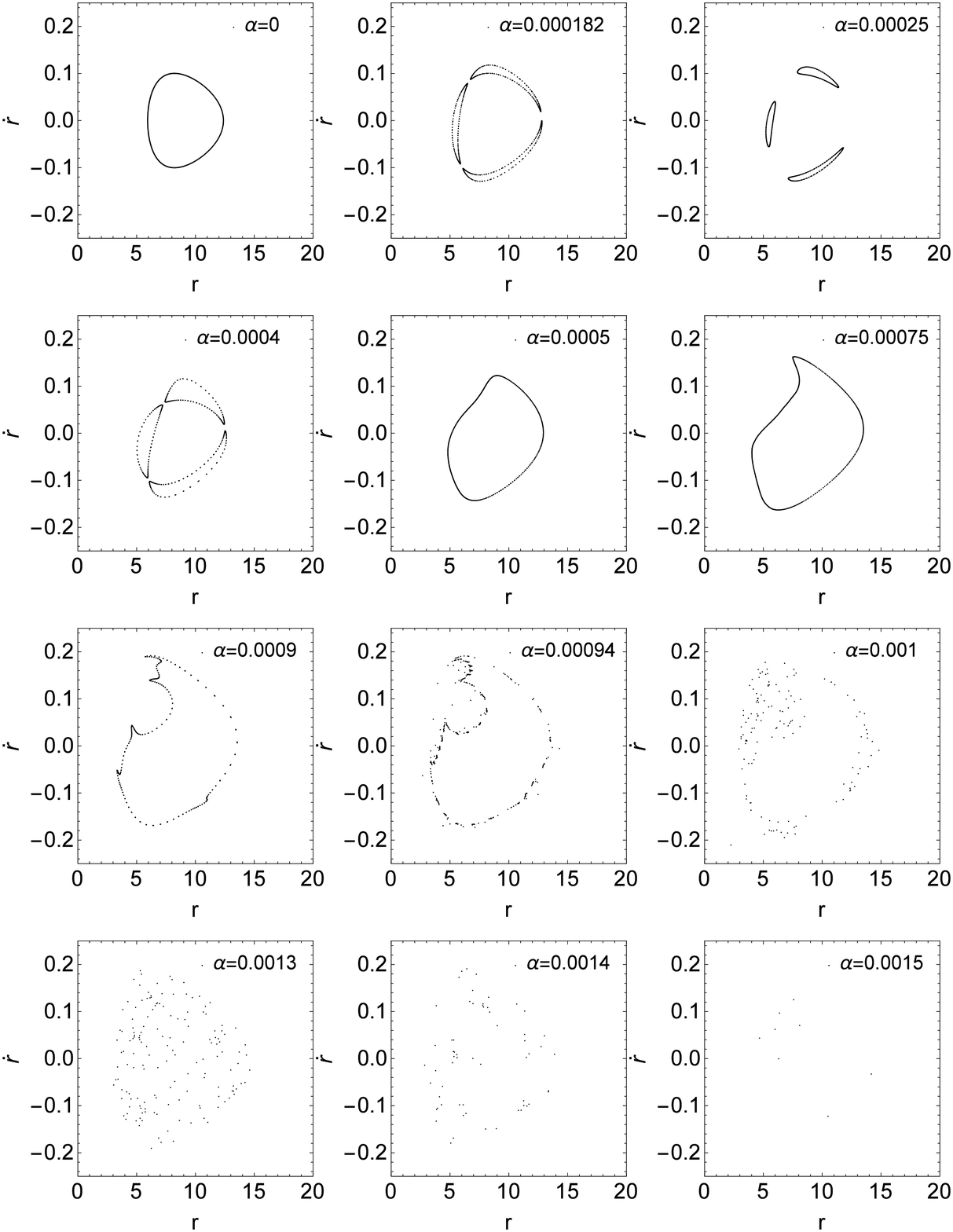}
\caption{ The change of
Poincar\'{e} section ($\theta=\frac{\pi}{2}$ on the plane $(r, \dot{r})$ ) with acceleration parameter $\alpha$ for the motion of the timelike particle in the accelerating and rotating  black hole spacetime with the fixed parameters $a=0.6$, $E=0.95$, $L=3M$ and the initial conditions $\{$ $r(0)=8$; $\dot{r}(0)=0.1$; $\theta(0)=\frac{\pi}{2}$$\}$. }
\end{figure}
\begin{figure}[ht]
\includegraphics[width=15cm]{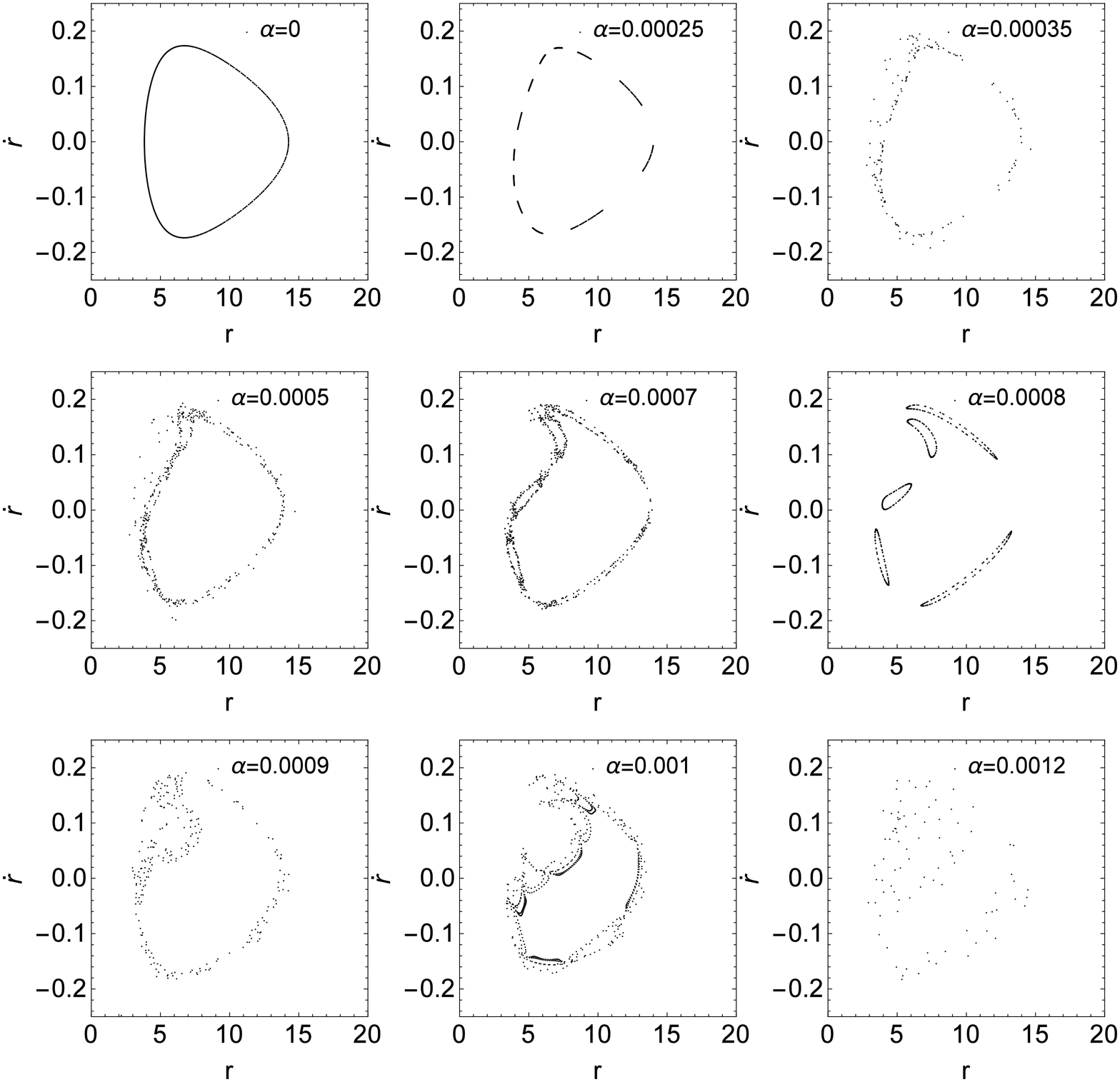}
\caption{ The change of
Poincar\'{e} section ($\theta=\frac{\pi}{2}$ on the plane $(r, \dot{r})$ ) with acceleration parameter $\alpha$ for the motion of the timelike particle in the accelerating and rotating  black hole spacetime with the fixed parameters $a=0.6$, $E=0.95$, $L=3M$ and the initial
conditions $\{$ $r(0)=12$; $\dot{r}(0)=0.1$; $\theta(0)=\frac{\pi}{2}$$\}$. }
\end{figure}

In Fig.(7), Poincar\'{e} sections with $\theta=\frac{\pi}{2}$ on the plane $(r, \dot{r})$  for different acceleration parameters $\alpha$ are plotted for the motion of a timelike particle in the accelerating
and rotating  black hole spacetime with the fixed parameters $a=0.6$, $E=0.95$, $L=3M$ and the initial
conditions $\{$ $r(0)=8$; $\dot{r}(0)=0.1$; $\theta(0)
=\frac{\pi}{2}$$\}$. We find that for $ \alpha\in [0,0.009]$ the phase path
is a quasi-periodic Kolmogorov-Arnold-Moser (KAM) tori and the behavior of this system is non-chaotic.  Especially,  there are more complicated KAM tori trajectories for $\alpha\in [0.000182,0.0004]$. It is composed of three  secondary KAM tori belonging to the same trajectories where the successive points jump from one loop to the next. These little loops are called a chain of islands. With the acceleration parameter $\alpha$ increasing, the chain of islands are joined together and become a big KAM tori. This shows that trajectory is regular and integrable in this case. However, when $\alpha$ is larger than $0.0009$, KAM tori is destroyed and the corresponding trajectory is non-integrable, which indicates that the behavior of this system is chaotic. Especially, for $ \alpha=0.001\sim 0.0013$, the tori is completely destroyed and the pattern is composed of discrete points, which means that the chaotic behavior becomes stronger  with the acceleration $\alpha$. From Fig.(7), we also note that there exist a few discrete points in the Poincar\'{e} section in the case with $\alpha=0.0015$, which is caused by a fact that the particle falls finally into the acceleration horizon of the black hole after undergoing several chaotic oscillations around black hole in this case. Thus, it is different essentially from those in the case of usual multiple-periodic motion. It is well known that the behavior of non-linear dynamical system depends on the choice of the initial conditions. In Fig.(8), we adopt another initial conditions $\{$ $r(0)=12$; $\dot{r}(0)=0.1$; $\theta(0)
=\frac{\pi}{2}$$\}$ and investigate the dependence of
Poincar\'{e} sections (with $\theta=\frac{\pi}{2}$) on the acceleration parameters $\alpha$. It is shown that with the increase of value of $\alpha$,
KAM tori in the phase plane is destroyed gradually to dispersed points at first, and then it is recovered slowly to a close curve. With further increasing of $\alpha$, the close KAM tori is completely destroyed again. This means that the behavior of the system undergoes a process from regular to chaotic then to regular, and finally to chaotic. Corresponding,  the non-integrability of the motion of particle in Fig.(8) first increases and then decreases, and finally increases with $\alpha$. Similarly, for the larger $\alpha$, we obtain only a kind of unstable escaped solutions for this chosen initial condition as in the previous discussion. Thus, the dependence of the non-integrability of the motion on the acceleration parameter $\alpha$ depends on the initial conditions and the parameters of system.

In Figs.(9) and (10), we also plot Poincar\'e section on the
plane $(r, \dot{r})$ for the motion of the timelike particle with different initial conditions in the background of accelerating and rotating  black hole spacetime with different
parameters. According to previous discussion,  we here set the acceleration parameter in the range $0\leq\alpha<0.0016$. From Fig.(9), we  find that for the fixed rotation parameter $a=0.6$, the chaotic region first increases and then decreases with the increase of the acceleration parameter $\alpha$. Moreover, for the case $\alpha=0$, we find that there are a series of close
curves in the Poincar\'e section, which means that there do not
exist chaotic orbits in the Kerr black hole spacetime, which is
consistent with the previous discussion.
Fig.(10) tells us that for the fixed acceleration parameter
$\alpha=0.001$, the chaotic region first increases and then
decreases with the increase of the rotation parameter $a$.
It is shown clearly in the Poincar\'{e} section that the numbers and positions of fixed points of the system change with the rotation parameter $a$.
\begin{figure}[ht]
\includegraphics{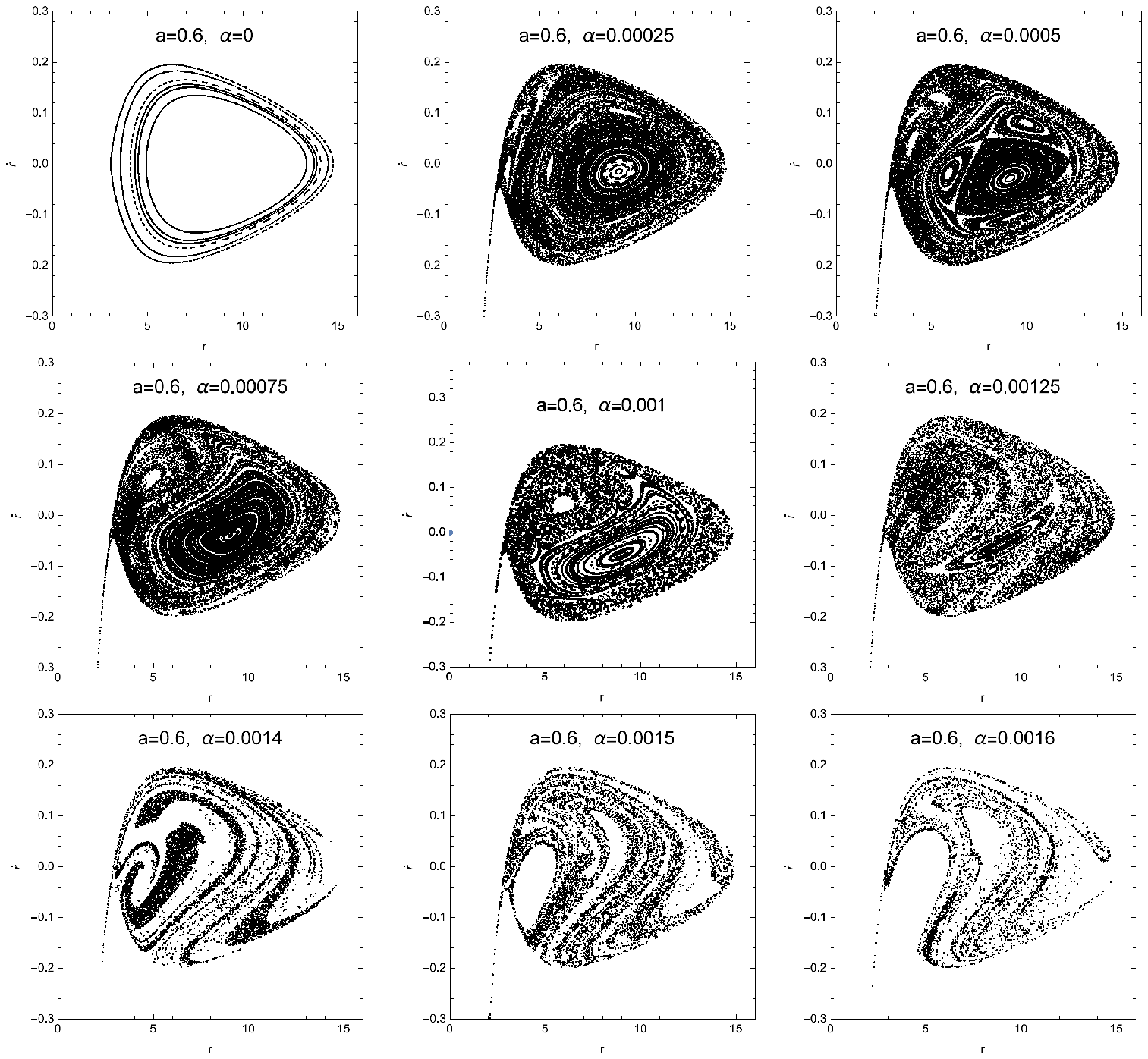}
\caption{ The change of Poincar\'{e} section ($\theta=\frac{\pi}{2}$ on the plane $(r, \dot{r})$ ) with the acceleration parameter $\alpha$ for the motion of the timelike particle in the accelerating
and rotating  black hole spacetime with the fixed parameters $a=0.6$, $E=0.95$ and $L=3M$.}
\end{figure}
\begin{figure}
\includegraphics{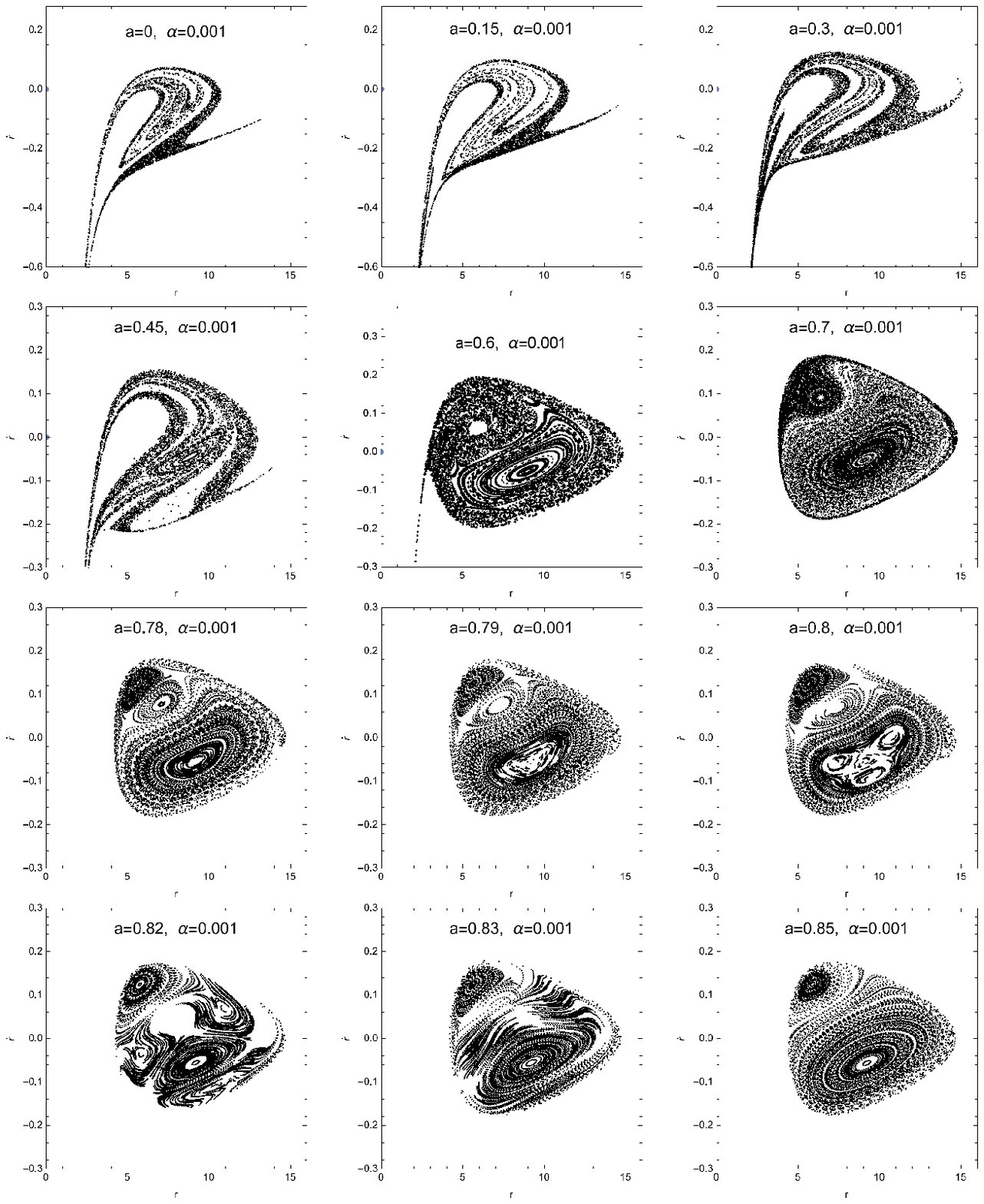}
\caption{ The change of
Poincar\'{e} section ($\theta=\frac{\pi}{2}$ on the plane $(r, \dot{r})$ ) with the rotation parameter $a$ for the motion of the timelike particle in the accelerating
and rotating  black hole spacetime with the fixed parameters $\alpha=0.001$, $E=0.95$ and $L=3M$.}
\end{figure}

The dependence of dynamical behaviors of system on the black hole parameters can also be visualised in the form of a bifurcation diagram. In Figs.(11) and (12), we plot the bifurcation diagram of the radial coordinate $r$ with the acceleration parameter $\alpha$ and the rotation parameter $a$ for the particle motion in the accelerating and rotating black hole spacetime with fixed $E=0.95$ and $L=3M$. Here we chose the set of initial conditions are $\{$ $r(0)=8$; $\dot{r}(0)=0$; $\theta(0) =\frac{\pi}{2}$$\}$. We find that for the case $\alpha=0$ there
is only a periodic solution and no bifurcation for the dynamical system (\ref{eqmotion}), which confirms again that the motions of particles are not chaotic in the rotation black hole spacetime without acceleration. For the case with non-zero acceleration, it is obvious that there exist periodic, chaotic and escaped solutions which depend on the acceleration parameter $\alpha$ and the rotation parameter $a$. For the chaotic solution, one can find
the range of $r$ in the bifurcation diagram increases almost with the acceleration parameter $\alpha$, which means that $\alpha$ enhances the strength of chaotic motion. With increase of the rotation parameter $a$, the range of $r$ in the chaos solution first decreases and then increases for the smaller $\alpha$, but it decreases for the larger $\alpha$.
Moreover, we find that the range of $a$ in which there exists escaped solution increases with $\alpha$. This could be explained by a fact that for the larger $\alpha$ the acceleration horizon $r_A$ is closer to the event horizon $r_H$ so that the particle is out-off-balance which yields that the particle falls either into
the event horizon or into the acceleration horizon of the black
hole in this case. These results show that the acceleration brings richer properties for the geodesic motion of particles in the accelerating and rotating black hole spacetime.
\begin{figure}
\includegraphics{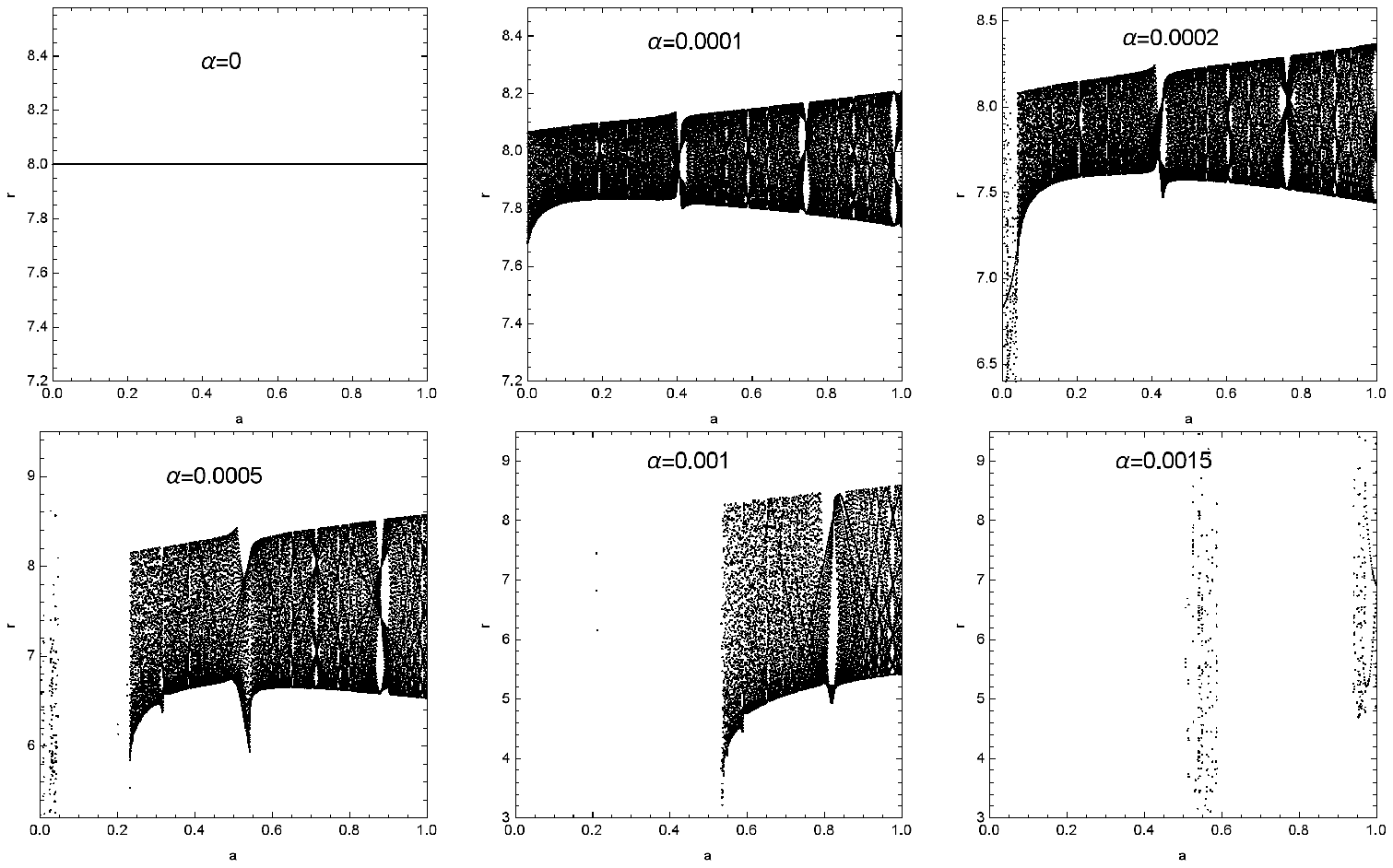}
\caption{ The bifurcation with the rotation parameter $a$ for the
motion of the timelike particle with the parameters $E=0.95$ and $L=3M$ in the accelerating and rotating
black hole spacetime. The set of initial conditions are $\{$ $r(0)=8$;
$\dot{r}(0)=0$; $\theta(0) =\frac{\pi}{2}$$\}$.}
\end{figure}
\begin{figure}
\includegraphics{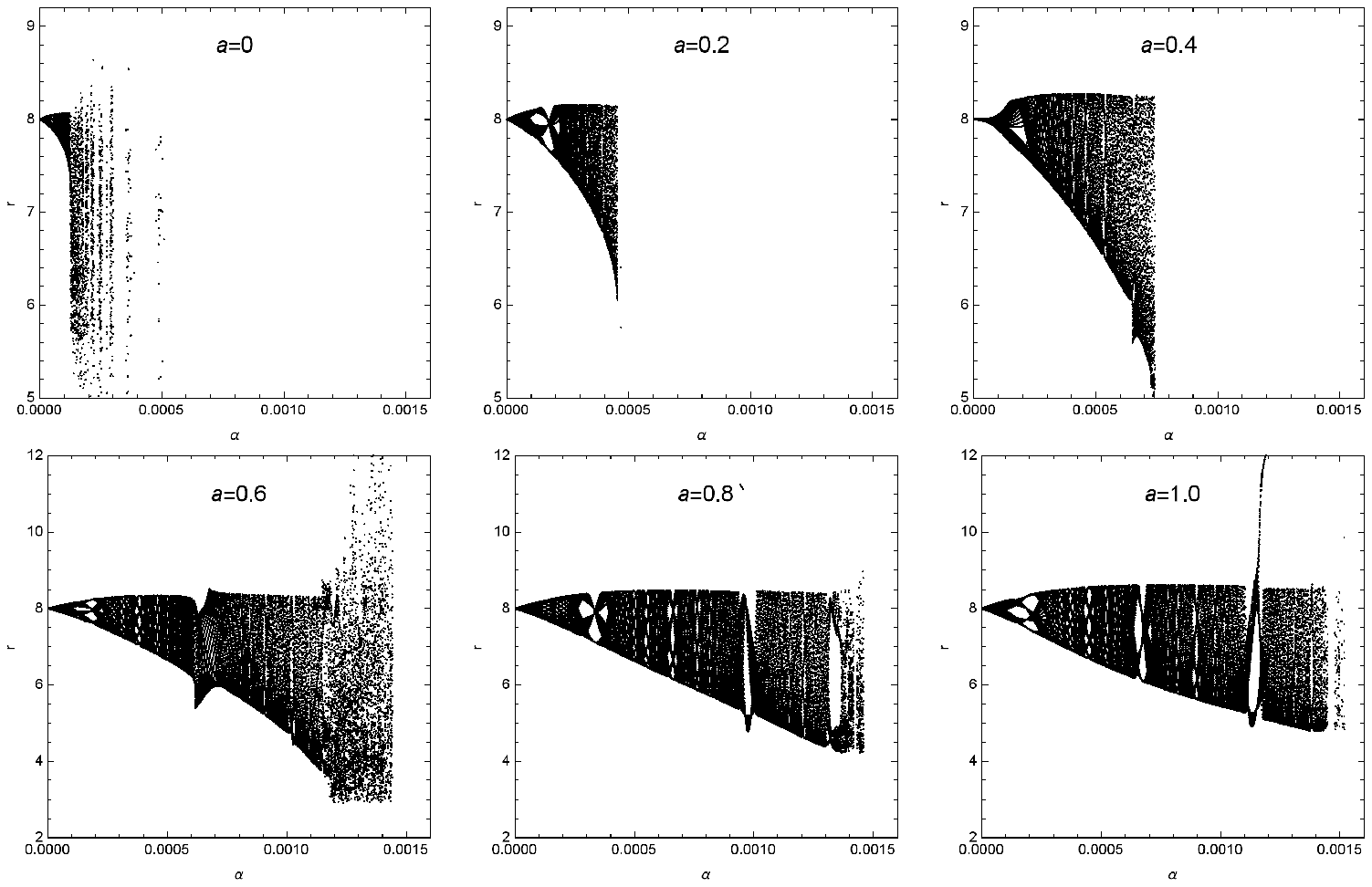}
\caption{ The bifurcation with the acceleration parameter $\alpha$
for the motion of the timelike particle with the parameters $E=0.95$ and $L=3M$ in the accelerating and
rotating black hole spacetime. The set of initial conditions are $\{$ $r(0)=8$;
$\dot{r}(0)=0$; $\theta(0) =\frac{\pi}{2}$$\}$.}
\end{figure}

\section{Summary}

In summary, we have studied the motion of timelike particles along geodesic in the background of
accelerating and rotating black hole spacetime by Poincar¨¦ sections, the frequency spectrum and the power spectrum, the fast Lyapunov exponent indicator, and the bifurcation diagram. Our results confirm that the chaos exists in the geodesic
motion of the particles. It is mainly because the presence of the acceleration parameter $\alpha$ yields that
the equations of motion are not be variable-separable and the corresponding dynamical system is non-integrable.
Moreover, we probe the effects of the acceleration and
rotation parameters of black hole on the chaotic behavior of a timelike geodesic particle. Our results show that the dependence of the non-integrability and the chaotic motion on the acceleration parameter $\alpha$ depends on the initial conditions and the parameters of system.
For the fixed acceleration parameter $\alpha=0.001$, we find that the chaotic region in Poincar¨¦ sections first increases and then decreases with the increase of the rotation parameter $a$.
For the fixed rotation parameter $a=0.6$, the chaotic region first increases and then decreases with $\alpha$. For the particle with chaotic motion, we find
the range of $r$ in the bifurcation diagram increases almost with the acceleration parameter $\alpha$, which means that $\alpha$ enhances the strength of chaotic motion. With increase of the rotation parameter $a$, the range of $r$ in the chaos solution first decreases and then increases for the smaller $\alpha$, but it decreases for the larger $\alpha$. When $\alpha=0$, we find that it can be reduced to the case of Kerr black hole spacetime in which there does not exists chaotic motion of particle.
Our results show that the acceleration yields richer effects on the geodesic motion of particles in the accelerating and rotating black hole spacetime.

\section{\bf Acknowledgments}

We would like to thank Prof. Wenhua Hai and Prof. Guishu Chong for their useful discussions of the early stage of this
work. This work was partially supported by the National Natural
Science Foundation of China under Grant No.11275065,  No. 11475061,
the construct program of the National Key Discipline, and the Open
Project Program of State Key Laboratory of Theoretical Physics,
Institute of Theoretical Physics, Chinese Academy of Sciences, China
(No.Y5KF161CJ1).

\end{document}